\documentstyle[10pt, aas2pp4]{article}
\slugcomment{Accepted for publication in ApJ}
\lefthead{KIM, MORRIS, \& LEE}
\righthead{EVAPORATION OF COMPACT YOUNG CLUSTERS}

\def\spose#1{\hbox to 0pt{#1\hss}}
\newcommand\lsim{\mathrel{\spose{\lower 3.0pt\hbox{$\mathchar"218$}}
     \raise 2.0pt\hbox{$\mathchar"13C$}}}
\newcommand\gsim{\mathrel{\spose{\lower 3.0pt\hbox{$\mathchar"218$}}
     \raise 2.0pt\hbox{$\mathchar"13E$}}}
\newcommand\msun{{\rm \,M_\odot}}

\begin{document}
\title{EVAPORATION OF COMPACT YOUNG CLUSTERS\\
NEAR THE GALACTIC CENTER}
\author{Sungsoo S. Kim}
\affil{Department of Physics, Korea Advanced Institute of Science \& Technology,
Daejon 305-701, Korea}
\authoremail{sskim@space.kaist.ac.kr}
\author{Mark Morris}
\affil{Department of Physics \& Astronomy, University of California,
Los Angeles, CA 90095-1562;\\
Institut d'Astrophysique de Paris, 98 bis Bvd. Arago, 75014 Paris, France}
\authoremail{morris@astro.ucla.edu}
\author{Hyung Mok Lee\altaffilmark{1}}
\affil{Department of Astronomy, Seoul National University, Seoul 151-742, Korea}
\authoremail{hmlee@astro.snu.ac.kr}

\altaffiltext{1}{This study was initiated when he was at
Pusan National University, Dept. of Earth Sciences, Korea.}

\vskip -1cm
\begin{abstract}
{\normalsize
We investigate the dynamical evolution of compact young clusters (CYCs)
near the Galactic center (GC) using Fokker-Planck models.  CYCs are
very young ($<5$~Myr), compact ($<1$~pc), and only a few tens of pc away
from the GC, while they appear to be as massive as the smallest Galactic
globular clusters ($\sim 10^4 \msun$).  A survey of cluster
lifetimes for various initial mass functions, cluster masses, and
galactocentric radii is presented.  Short relaxation times due to the
compactness of CYCs, and the strong tidal fields near the GC
make clusters evaporate fairly quickly.  Depending on cluster
parameters, mass segregation may occur on a time scale shorter than
the lifetimes of most massive stars, which accelerates the cluster's
dynamical evolution even more.
When the difference between the upper and lower mass
boundaries of the initial mass function is large enough,
strongly selective ejection of lighter stars makes massive stars dominate
even in the outer regions of the cluster, so the dynamical evolution of
those clusters is weakly dependent on the lower mass boundary.
The mass bins for Fokker-Planck simulations were carefully chosen to properly
account for a relatively small number of the most massive stars.
We find that clusters with a mass $\lsim 2 \times 10^4 \msun$
evaporate in $\lsim 10$~Myr.  Two CYCs observed near the GC
--- the ``Arches cluster" (G0.121+0.17) and the ``Quintuplet cluster"
(AFGL2004) --- are interpreted in terms of the models; their central
densities and apparent ages are consistent with the hypothesis that they
represent successive stages of cluster evolution along a common track,
with both undergoing rapid evaporation.   A simple calculation
based on the total masses in observed CYCs and the lifetimes obtained here
indicates that the massive CYCs comprise only a fraction of the star
formation rate (SFR) in the inner bulge estimated from Lyman continuum
photons and far-IR observations.
This is consistent with the observation that many stars
in the inner bulge form outside the large clusters.
}
\end{abstract}


\vskip 0.6cm
\centerline{\large \it Accepted for publication in ApJ}
\vskip 0.6cm

\section{INTRODUCTION}
\label{sec:introduction}

The inner few hundred pc of the Galactic bulge (the inner bulge)
contains stars of a variety of ages, in addition to an apparently coeval
population of galactic-age stars.  The evidence
for recent massive star formation there
has grown with the observations of several
clusters of emission-line stars in that region.  Intermediate-age stars
have also been observed in the inner bulge in the form of OH/IR
stars (\markcite{LHW92}Lindqvist, Habing, \& Winnberg 1992;
\markcite{Se98}Sjouwerman et al. 1998).
Noting the presence of these young and intermediate-age stars,
\markcite{SM95}Serabyn \& Morris (1995) argued that star
formation in the central molecular zone occupying
the inner bulge has been sustained over the lifetime of the
Galaxy in spite of the inhospitality of that environment for star formation,
owing to large gas temperatures and turbulent velocities, strong magnetic
fields, and strong tidal forces.

About half a dozen distinct sites of recent star formation have been
found in the inner bulge, most notably including the ``Arches cluster"
(G0.121+0.017;
\markcite{Ne95}Nagata et al. 1995; \markcite{Ce96}Cotera et al. 1996;
\markcite{SSF98}Serabyn, Shupe, \& Figer 1998) and the ``Quintuplet cluster"
(AFGL2004; \markcite{Oe90}Okuda et al. 1990; \markcite{Ne90}Nagata et al.
1990; \markcite{GMM90}Glass, Moneti, \& Moorwood 1990;
\markcite{FMM95}Figer, McLean, \& Morris 1995, \markcite{FMM99}1999),
both lying within $\sim 35$~pc, in projection, of the Galactic center (GC).
These clusters are very young ($<5$~Myr), and compact ($<1$~pc), while
they appear to be as massive as the smallest Galactic globular clusters
($\sim 10^4
\msun$).  Their young ages are manifested by the presence of ample massive
stars:  the Arches and Quintuplet contain about 120 and 30 stars
having initial mass larger than $20 \msun$, respectively
(\markcite{SSM98}Serabyn et al. 1998; \markcite{FMM99}Figer et al. 1999).
\markcite{M93}Morris (1993) suggested that the non-standard star formation
environment near the GC may lead to an initial mass
function (IMF) skewed toward relatively massive stars (flatter IMF)
and having an elevated mass cutoff.  However, the observational limit for
the lowest mass in these clusters is still as high as several $M_\odot$,
so not much is known about the lower end of their mass function.

The Arches and Quintuplet clusters are two members of what appears
to be a special category of clusters; indeed, their large masses
place them at the lower end of the category of super star clusters
(e.g., \markcite{HF96}Ho \& Filippenko 1996).
Another known, potential member of this category near the GC
is the young cluster in the central parsec, but because
of its unusual location at the bottom of the Galactic gravitational
potential well, and in the immediate vicinity of the central
supermassive black hole, it may have had quite a different origin.
Thus, it is not clear that this cluster should be categorized
alongside the Arches and Quintuplet, even though it has a
comparable mass (\markcite{FMM99}Figer et al. 1999).  In any case, the
calculations presented in this paper do not apply to the
central cluster.  The stellar cluster in Sgr B2 is potentially
another supercluster currently in formation
(e.g., \markcite{Ge95}Gaume et al. 1995),
although it is not yet clear whether it will attain the status
of the two more evolved clusters.

With the observations of these compact, young clusters (CYCs) near the
GC, a natural question arises:  Why don't we observe older examples
of CYCs?  Are we witnessing very unusual events that happened to have
recently taken place in the GC, or are older clusters rare because they
evaporate in a relatively short period of time?
The proximity of the CYCs to the GC implies a strong tidal field,
and the short relaxation time due to their compactness makes very
massive stars play an important dynamical role through mass segregation,
which accelerates the dynamical evolution even more, before these stars
disappear with mass loss and explosions.
These two characteristics suggest fairly short lifetimes for the CYCs.
The ages of the Arches and Quintuplet clusters are estimated to
differ by about a factor of two, providing an excellent opportunity
to compare observations with the numerical study of their dynamical
evolution.  Because their rather extreme parameters have only recently
been recognized, these clusters have not yet been numerically and
theoretically explored.
In the present study,
we investigate the clusters' lifetime, $t_{ev}$, for various
parameters plausible for CYCs, and compare the results with current
observations of CYCs and with estimates of the star formation rate
in the inner bulge.

A discussion of the time scales related to the dynamical evolution
of spherical stellar systems and on our choice of simulation methods
is given in \S~\ref{sec:backgrounds}.  We describe the details of
our models in \S~\ref{sec:models}, and show the simulation
results in \S~\ref{sec:results}.  The implications of our results are
discussed in \S~\ref{sec:discussion}, and \S~\ref{sec:summary} summarizes
our findings.

\section{BACKGROUNDS}
\label{sec:backgrounds}

\subsection{Time Scales}
\label{sec:timescales}

The dynamical evolution of stellar systems is affected by many factors such
as galactic tidal fields, the stellar mass function, and stellar evolution,
as well as their overall mass and physical extent.  In general, a cluster loses
its mass by two-body relaxation and stellar evolution (via supernovae
and stellar winds), and the resulting dynamical evolution of the cluster is
accelerated by the presence of a strong tidal field.

The evolution of star clusters in a static tidal field has been
studied extensively. \markcite{H61}H\'enon (1961) obtained a self-similar
solution for an expanding cluster assuming that there is a central
energy source (which he speculated to be binaries). Such a cluster
is found to evaporate completely in 22.5 times the half-mass
relaxation time, which is defined as
\begin{equation}
\label{trh}
t_{rh} = 0.138 {M^{1/2} R_h^{3/2} \over \bar m G^{1/2}
                        \ln \Lambda},
\end{equation}
where $M$ is the total mass of the cluster, $R_h$ the half-mass radius,
${\bar m}$ the mean mass of each star, and $G$ the gravitational constant.
The $\ln\Lambda$ term appears because of the finite size of the cluster,
and is called the Coulomb logarithm.

The numerical integration of the Fokker-Planck equation by
\markcite{LO87}Lee \& Ostriker (1987)
confirmed H\'enon's general result. These works have been extended to
multi-mass models for typical globular cluster parameters
by \markcite{LFR91}Lee, Fahlman, \& Richer (1991)
and \markcite{LG95}Lee \& Goodman (1995), and the
lifetimes of the clusters in units of half-mass relaxation time have been
found to be shorter than those of single mass models by about a significant
factor.  Thus, clusters with a realistic mass function would survive less
than 10 $t_{rh}$.


The evaporation of stars from the cluster is due to relaxation, but
it takes some time for a star to escape from the cluster.  Since this
``lingering time" is of order of an orbital period of the star, the
evaporation of a cluster depends on both relaxation time and
dynamical time.  The average orbital time scale for stars in the cluster
can be expressed as:
\begin{equation}
\label{td}
	t_{dyn} \approx \left ( {R_t^3 \over G M} \right )^{1/2},
\end{equation}
where $R_t$ is the tidal radius of the cluster,
\begin{equation}
\label{rt}
        R_t =\left({M\over 2M_g}\right)^{1/3} R_g.
\end{equation}
Here, $R_g$ is the galactocentric radius, and $M_g$ the enclosed galactic
mass within $R_g$.  Equations (\ref{trh}) and (\ref{td}) give a relation
$t_{dyn} / t_{rh} \propto N^{-1}$, where $N$ is the number of stars.  Thus
the evaporation depends more sensitively on $N$ when $N$ is small.
Note that since $M_g$ is a function of $R_g$ (see eq. [\ref{mg}]),
equations (\ref{td}) and (\ref{rt}) imply that $t_{dyn}$ can be expressed
in terms of $R_g$ only.

Stellar evolution affects a cluster's dynamical evolution in several ways.
First, mass loss by stellar evolution
(through stellar winds and explosions) itself accelerates the mass loss
of a cluster and thus shortens the cluster's lifetime.  Furthermore,
the mass loss by stellar evolution, together with tidal evaporation,
may prevent the core collapse.
Since stars bound to a cluster have negative energy and
massive stars are predominantly located in the core due to mass segregation,
the mass loss from the cluster thus has an indirect heating effect.

The effect of very massive stars may be neglected for globular clusters,
which have initial $t_{rh}$ of $\sim 10^{9-10}$~yr, much
longer than evolutionary time scales of high mass stars.
The low mass stars eventually undergo evolution on a much longer time
scale, and their effects on dynamics should be very small.
Therefore, in old globular clusters, the stellar evolution should have
been effective only in the initial phase when the relaxation did not
play any role. The subsequent evolution should have been governed
mostly by the relaxation process alone.
Thus the study of globular cluster evolution may assume
that clusters start with upper mass boundary, $m_u$, much lower than the real
initial value (for example, for $m_u$, \markcite{CW90}Chernoff \& Weinberg
1990 adopted $15 \msun$ with stellar evolution, and \markcite{LFR93}Lee,
Fahlman, \& Richer 1991 adopted $0.8 \msun$ without stellar evolution).
However, the sizes of CYCs near the GC
are very small for their masses, leading to initial $t_{rh}$
as short as $\sim 10^{6-7}$~yr.  Moreover,
the mass segregation time, $t_{rh} \bar m / m_u$, may be even shorter
than the lifetime of the most massive star.  Then very massive stars
play an important role in the evolution of the cluster through
rapid mass segregation and subsequent demise.

To summarize, the dynamical evolution of a cluster is determined
by relative differences
between time scales for the above processes, two-body relaxation including
core collapse and subsequent expansion, tidal evaporation, stellar evolution,
and mass segregation.  For example, stellar evolution determines the
cluster evolution time scale when the chemical evolution of massive
stars occurs much faster than the relaxation time scale of the cluster,
and the relaxation governs the cluster evolution when its time scale
is shorter than that of stellar evolution.  Also, if a cluster has a large
stellar mass range, the segregation of massive stars toward the central parts
shortens the relaxation
time in the core and makes relaxation processes more important.  Finally,
a strong tidal field accelerates all these processes.
Since the time scales of these processes
scale with cluster parameters in different ways, scaling
the evolutionary route of one cluster to another is quite limited,
and numerical simulations are generally required for the unexplored parameter
regime.

\subsection{N-body vs. Fokker-Planck Models}

For the study of dynamical evolution of spherical stellar systems,
N-body and Fokker-Planck simulations are among the most widely used
methods.  The N-body method is certainly more realistic and requires
less assumptions than other methods, but real-number N-body simulations
for CYCs having $N \sim 10^{3-5}$ are still prohibitively expensive
in terms of computing time, especially in the context of a simulation survey
in which a range of parameters is being explored.  Many studies based on
the N-body method use a time-scaling technique to simulate a large-number
system with a smaller number of stars on behalf of computing cost.
However, correct scaling is not always guaranteed when the stellar
evolution time scale, $t_{se}$, is not well separated from $t_{dyn}$
and $t_{rh}$, as might be the case of CYCs (see, however,
\markcite{AH98}Aarseth \& Heggie 1998 for a compromise technique,
``variable scaling'').  Also, as mentioned above, the evaporation
rate of tidally-limited clusters is a function of both
$t_{rh}$ and $t_{dyn}$, which scale differently with $N$ and $M$.
Thus, direct N-body calculations of the system with a smaller number
of stars are not able to
correctly mimic the evaporation of the system with a larger N.
Moreover, when the mass function is steep, the upper mass boundary, $m_u$,
can be a function of the total number of stars in the cluster
for statistical reasons (the more stars, the higher $m_u$).
Then the system with a smaller number could not properly represent
the full mass range of the larger system.

On the other hand,
although the Fokker-Planck method requires more assumptions and
approximations, it is much less expensive and gives statistically
correct results.  These features are critically beneficial,
not only when the study is meant for a parameter survey, but also when
statistically stable results without random noise are necessary
for theoretical analyses.  For these reasons, as a first step of
our study of the fate of CYCs, we simulate the dynamical evolution
of CYCs using Fokker-Planck models.  We are planning N-body simulations
for a few representative cluster parameter sets in a followup study.

A recent comparison between Fokker-Planck and N-body simulations
for tidally-limited clusters showed a discrepancy in cluster lifetime
between the two methods
(\markcite{FH95}Fukushige \& Heggie 1995; \markcite{PZe98}Portegies Zwart
et al. 1998).  However, \markcite{TPZ98}Takahashi \& Portegies Zwart (1998)
were able to show that this discrepancy can be successfully removed
by adopting anisotropic Fokker-Planck models with the ``apocenter
criterion'' and an appropriate constant for the speed of star removal
behind the tidal radius.  Here we adopt anisotropic Fokker-Planck
models and the apocenter criterion as an effort to minimize the
possible discrepancy with N-body simulations.

\section{MODELS}
\label{sec:models}

We have used the two-dimensional (energy-angular momentum space),
orbit-averaged Fokker-Planck code by \markcite{T97}Takahashi (1997).
The code assumes that star clusters are spherically symmetric and are in
dynamical equilibrium.

For our initial models, we adopt a multi-mass distribution function
composed of single mass King models with equal velocity dispersions.
It is assumed that clusters initially fill the tidal radius and that
all stars in the cluster are formed simultaneously at $t=0$.  The
relation between $M$ and $R_t$ is determined via equation~(\ref{rt}) and
\begin{equation}
\label{mg}
        M_g = 2 \times 10^8 \msun \, \left ( {R_g \over {\rm 30 \, pc}}
                \right )^{1.2},
\end{equation}
which was adopted from \markcite{GT87}Genzel \& Townes (1987).
Primordial binaries are not considered in this study, and the
Coulomb logarithm is taken as $\ln \Lambda = \ln N$.

Heating by binaries formed through three-body processes and tidal-capture
processes is included.  For the heating rate per unit volume by three-body
binaries, we adopt the formulation by \markcite{LFR91}Lee et al. (1991):
\begin{equation}
\label{E3b}
        \dot E_{3b} = 4.21 \times 10^3 \, G^5 \left ( \sum_i {n_i m_i^2 \over
                        v_i^3} \right )^3 v_c^2,
\end{equation}
where the summation is over all mass components,
$n_i$ and $v_i^2$ are the number density and velocity
dispersion of component $i$, respectively, and $v_c^2$ is the mass-weighted,
central velocity dispersion.  On the other hand,
we use a modified version of the tidal-capture binary heating rate per
unit volume by \markcite{LO93}Lee \& Ostriker (1993):
\begin{equation}
\label{Etc}
        \dot E_{tc} = \sum_i \sum_{j \ge i} (m_i+m_j+\bar m_c) \sigma_{tc}
                        v_{ij} n_i n_j \phi_c,
\end{equation}
where the summation $i$ is only for main-sequence stars, $\bar m_c$ is the mean
mass in the center, $\sigma_{tc}$ the tidal-capture cross section, $v_{ij}$
the relative rms velocity between components $i$ and $j$, and $\phi_c$
the central gravitational potential.  The fitting formulae of
\markcite{KL99}Kim \& Lee (1999) were adopted for $\sigma_{tc}$.

The effect of the Galactic tidal field is assumed to be constant
(circular orbit around the GC and spherically symmetric
potential due to the tidal force) and is realized by imposing a tidal
boundary.  A star is removed if its apocenter distance, $R_a$, exceeds
the tidal radius, $R_t$ (apocenter criterion; \markcite{TLI97}Takahashi, Lee,
\& Inagaki 1997).  Removal of stars satisfying the apocenter criterion
is realized by the formalism of \markcite{LO87}Lee \& Ostriker (1987):
\begin{equation}
        {df \over dt} = - \alpha_{esc} f \left [ 1- \left ( {E \over E_t}
                        \right )^3 \right ]^{1/2} {1 \over 2 \pi} \sqrt{
                        {4 \pi \over 3} G \rho_t},
\end{equation}
where $f$ is the distribution function, $E_t$ the tidal energy, and
$\rho_t$ the mean mass density within the tidal radius.  This formula
accounts for the persistence time of an escaping star, and the speed of
star removal is determined by a dimensionless constant $\alpha_{esc}$.
Following \markcite{TPZ98}Takahashi \& Portegies Zwart (1998), we
have adopted $\alpha_{esc} = 2$.

\begin{figure}
\centerline{\epsfxsize=8.8cm\epsfbox{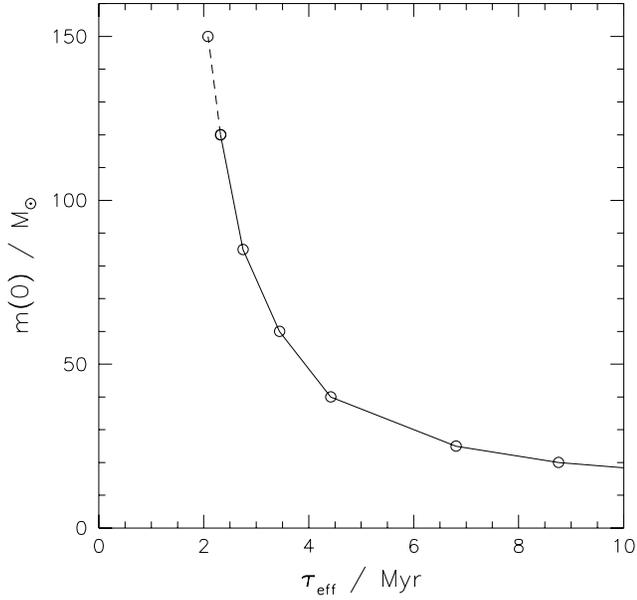}}
\caption
{\label{fig:eff}The effective lifetime $\tau_{eff}$ of a star with
initial mass $m(0)$ defined by equation (\ref{taueff}).
Schaller et al. (1992) was adopted for $m(t)$.
A logarithmic extrapolation has been made for $m(0)>120 \msun$.
These $\tau_{eff}$ values are used for $\tau_{i \pm {1 \over 2}}$ in
equation~(\ref{mevol}).
}
\end{figure}

The effect of stellar evolution is realized by secularly decreasing the mass of
each component.  Following \markcite{CW90}Chernoff \& Weinberg (1990),
the mass of each component is linearly decreased by
\begin{eqnarray}
\label{mevol}
        m_i(t) = m_i(0) - [m_i(0)-m_{i,f}] {(t- \tau_{i+{1 \over 2}})
                        \over (\tau_{i-{1 \over 2}}-\tau_{i+{1 \over 2}})}
			\nonumber \\
                        \hfill {\rm for}\;\;\; \tau_{i+{1 \over 2}} < t <
                        \tau_{i-{1 \over 2}},
\end{eqnarray}
where $m_i(t)$ is the mass of component $i$ at time $t$, $m_{i,f}$ the final
mass of the component, $\tau_{i \pm {1 \over 2}}$ the lifetime of a star
with mass $m_{i \pm {1 \over 2}}$ (see below for $m_{i \pm {1 \over 2}}$).
For $t < \tau_{i+{1 \over 2}}$,
$m_i(t)$ has a constant value of $m_i(0)$, and for $t > \tau_{i-{1 \over 2}}$,
$m_i(t)$ is set to $m_{i,f}$.  For intermediate and low mass stars, the
mass evolution between the main-sequence and degenerate stage
is so abrupt that it can be approximated with a step function.  However,
in case of massive stars ($m > 20 \msun$), the mass evolution is
continuous throughout the stellar lifetime, so it is quite difficult to define
the time for which a star maintains its initial mass and after which
the star becomes degenerate.  Furthermore, it is not trivial to implement
such continuous mass loss into a mass component which is a representation
of a set of stars with a certain range of masses and mass loss rates.
For this reason, here we define an ``effective lifetime'' of a star such that
\begin{equation}
\label{taueff}
        \tau_{eff} = {1 \over m(0)} \int_0^{t_f} m(t) \, dt,
\end{equation}
where $m(t)$ is the mass evolution of a star with initial mass $m(0)$,
$t_f$ is the time required for the mass of a star
with initial mass $m(0)$ to become the final mass $m_f$.  We have
adopted \markcite{Se92}Schaller et al. (1992) for $m(t)$ and
$t_f$, and \markcite{D95}Drukier (1995) for $m_f$.  Figure \ref{fig:eff} is
a plot of $\tau_{eff}$ as a function of $m(0)$ obtained in this way.
Then this $\tau_{eff}$ is used for $\tau_{i \pm {1 \over 2}}$ in
equation~(\ref{mevol}).

The mass spectrum is represented by a set of discrete mass components.
For the mass binning, equal logarithmic space binning has been used
in many previous studies.  However, the same logarithmic binning
is not appropriate for representing largely different mass spectrum
slopes at the same time.  In particular, the logarithmic binning may
result in too small number of stars for the most massive bins of
a cluster with steep mass spectrum and/or with small $M$, which
is not desirable for Fokker-Planck models.  One could lower the upper
mass boundary in accordance with the cluster parameters
to avoid such a problem, but then the problem becomes
how one ``decides'' the upper mass boundary.  Generally, one is
interested in a cluster with given $M$, $\alpha$, and $m_l$.
In cases where the relaxation time of a cluster is much longer than
the stellar evolution time for massive stars, the effect of those
massive stars on the dynamical evolution of the cluster may be
neglected and the choice of $m_u$ is relatively unimportant.  However, in case
of our target clusters, the role of massive stars is expected to be
important because of their short relaxation times, and thus $m_u$
should be chosen carefully.  For this reason, we bin the mass
in a way that guarantees all mass components to have the number of stars
larger than a certain minimum value.  The mass range is binned to
make all components have equal value of
\begin{equation}
\label{mboundary}
        \int_{m_{i-{1 \over 2}}}^{m_{i+{1 \over 2}}} m^{\beta} N(m) \, dm,
\end{equation}
where [$m_{i-{1 \over 2}}$,$m_{i+{1 \over 2}}$] is the mass range of
a component $i$.  Thus the mass ranges of each component differ for
models with different parameters.
For a given $M$, $m_u$, $m_l$, $\alpha$, and $N_{bin}$ (number of bins),
one can determine the value of $\beta$ that assigns a stellar number of
$N_u$ to the most massive bin.  Once $\beta$ is determined,
$m_{i \pm {1 \over 2}}$ is obtained by equation~(\ref{mboundary})
with given $m_l$ and $m_u$.  Here we adopted $m_u = 150 \msun$,
$m_l = 0.1$,~$1 \msun$, $N_{bin}=15$, and $N_u=50$.
In this way, the effects of the most massive stars can be realized
in Fokker-Planck models correctly and consistently.
Note that $\beta=1$ is for equal mass binning, and $\beta=0$
is for equal number binning.

We choose the initial mass spectrum to be a simple power law
\begin{equation}
        dN(m) \propto m^{-\alpha} \, dm,
\end{equation}
where $dN(m)$ is the initial number of stars with masses between $m$
and $m+dm$, and $\alpha=2.35$ gives a Salpeter initial mass function.
The initial mass of component $i$, $m_i(0)$, is calculated by
\begin{equation}
        m_i(0) = { \int_{m_{i-{1 \over 2}}}^{m_{i+{1 \over 2}}}m^{1-\alpha}
                \, dm \over \int_{m_{i-{1 \over 2}}}^{m_{i+{1 \over 2}}}
                m^{-\alpha} \, dm} \, ,
\end{equation}
and the initial total mass of component $i$ is
\begin{equation}
        M_i(0) = \int_{m_{i-{1 \over 2}}}^{m_{i+{1 \over 2}}} N(m)m \, dm
\end{equation}
The number of stars in component $i$, $N_i$, is then defined to be $M_i/m_i$.
In Table \ref{table:runs}, $\beta$ and the initial mass of the most massive
component, $m_{15}(0)$, are given for all models.  Note that
$\beta$ and $m_{15}$ are determined by $M$, $m_l$, and $\alpha$
because we fix $m_u$, $N_{bin}$, and $N_u$.

\section{RESULTS}
\label{sec:results}

An example of the cluster evolution in our models is shown in
Figure~\ref{fig:evol}.
A mild core collapse takes place at $\sim 0.6$~Myr, and the postcollapse
expansion is accelerated by indirect heating due to stellar evolution starting
from $\sim 2.2$~Myr, which is clearly manifested by a rather abrupt
change in $R_t$ slope.  The mildness of the core collapse is only an apparent
phenomenon:  the relative increase of $\rho_c$ during the collapse becomes
larger when $\rho_c$ is plotted for only the most massive component.
While the core collapse and postcollapse
expansion time scales are determined by the size, structure, and mass function
of the cluster, the stellar evolution time scale is dependent only on
the mass function.  Since these time scales are comparable for most of our
models, the lifetime of a cluster will be determined jointly by these time
scales in a complex way.  In this section, we discuss the effect of
each cluster parameter on the evolution of the cluster. We pay
special attention to $t_{ev}$, which is given in Table~\ref{table:runs},
since this is closely related to the fate of the cluster.

Most of the discussion of evaporation and dynamical evolution assumes
that the time is expressed in units of relaxation time. That is useful if
the relevant process is related mostly to the two-body relaxation time.
However, the stellar evolution is important in the situation that we are
considering here, as discussed in \S~\ref{sec:timescales}.
Therefore, it is more convenient to express the results in absolute
time for the CYCs.

\subsection{Initial Concentration ($W_0$)}
\label{sec:w0}

\begin{figure}
\centerline{\epsfxsize=8.8cm\epsfbox{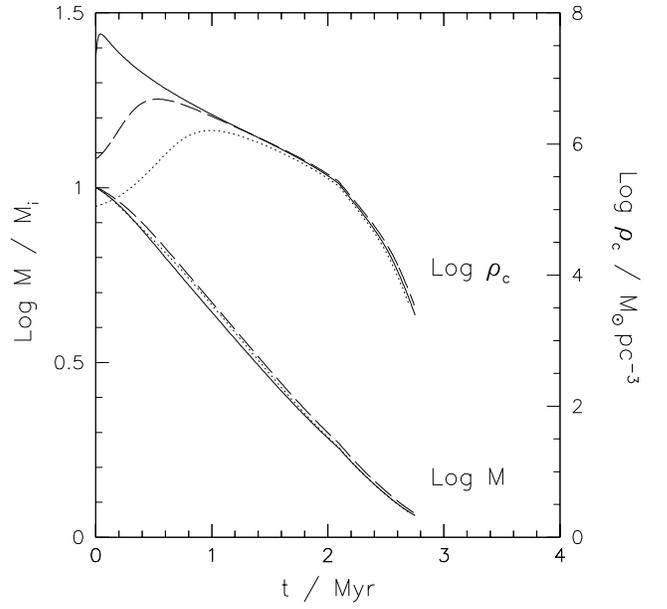}}
\caption
{\label{fig:w0}Evolution of $M$ and $\rho_c$ for models 111 ($W_0=1$; dotted
lines), 101 ($W_0=4$; dashed lines), and 112 ($W_0=7$; solid lines),
which have the same initial conditions except for $W_0$.  These models have
lagely different $t_{cc}$ but very similar $t_{ev}$.  This shows that
$t_{ev}$ is independent of $W_0$.
}
\end{figure}

King models, which we use as initial models, are a one-parameter family,
whose parameter $W_0$ determines the initial degree of central concentration
(the larger the value of $W_0$, the higher the concentration).
Figure \ref{fig:w0} shows the evolution of models 101, 111, and 112, which have
the same initial conditions except for  $W_0$.
While the times to total evaporation $t_{ev}$ are almost the same,
the epochs of the central density peak for these
three models are largely different.  The time to core collapse, $t_{cc}$,
is determined by the central relaxation time, $t_{rc}$, which is smaller
for more concentrated clusters, while the tidal evaporation
rate is determined by a cluster's global properties.  Clusters slowly expand
in the postcollapse phase, and this expansion accelerates the tidal
evaporation.
The evaporation times are nearly the same for these models
because we required our models to have the same $t_{dyn}$ (the same
$R_g$), $M$, and $R_t$.
The evaporation is not sensitive to the detailed structure of the cluster, but
depends mainly on macroscopic parameters such as $M$ and $R_{t}$.
Thus the insensitivity of $t_{ev}$ on $W_0$ will hold for the parameter
regime covered by our other models as well.
The evolution of the central parameters depends on $W_0$ just because
the core collapse is determined by the local conditions in the center, which
depend on the initial models.

\subsection{Mass Range ($m_l$)}
\label{sec:mrange}

\begin{figure}
\centerline{\epsfxsize=8.8cm\epsfbox{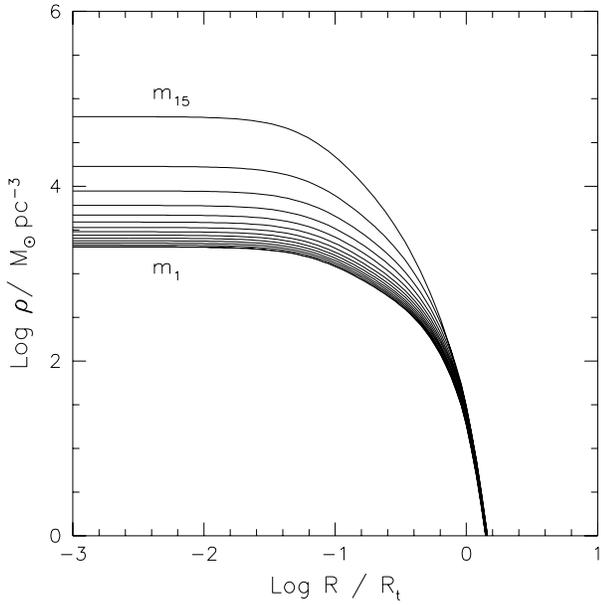}}
\caption
{\label{fig:prof}Density profiles of mass component 1 to 15 (bottom to top)
of model 142 at $t=2$~Myr (postcollapse phase).  Both core and envelope are
dominated by heavy components.
}
\end{figure}

The effects of lower and upper mass boundaries on $t_{ev}$ might be seen
by comparing models 101, 102, and 103.  One may expect that $m_l$ is
more important in determining $t_{ev}$ than $m_u$ because $\bar m$
varies more with $m_l$ than $m_u$ for our $\alpha$ values.  However,
interestingly, the results of models 101 to 103 show the opposite behavior:
$t_{ev}$ is approximately proportional to $m_{15}^{-1}$, but is almost
insensitive to $m_l$ (see next subsection for the relation between
$t_{ev}$ and $m_{15}$).  Generally, a cluster with a mass spectrum
rearranges itself so that heavies dominate the central region of the
cluster and lights the outer region in a time scale $\sim t_{rh}m_l/m_u$
($\sim t_{rh}m_l/m_{15}$ in our case; $m_{15}$ better represents the
upper mass boundary than $m_u$ does because $m_{15}$ accounts for the mass
spectrum; see \S \ref{sec:models}).  However, the mass ranges of the
above models are so wide (thus the ratios $m_{15}/m_l$ are large)
that even the outer region is dominated by heavies.
This can be clearly seen in the density
profile plot (Figure \ref{fig:prof}), where heavier components are
dominant throughout most of the cluster.  This presumably results from
the selective evaporation of lighter components due to equipartition.
However, the dominance of heavies in the outer
region does not solely explain the similar $t_{ev}$ of models 101
and 102 because the masses of the light components of these
models are largely different
and may contribute to $t_{rh}$ in dissimilar ways.

\begin{figure}
\centerline{\epsfxsize=8.8cm\epsfbox{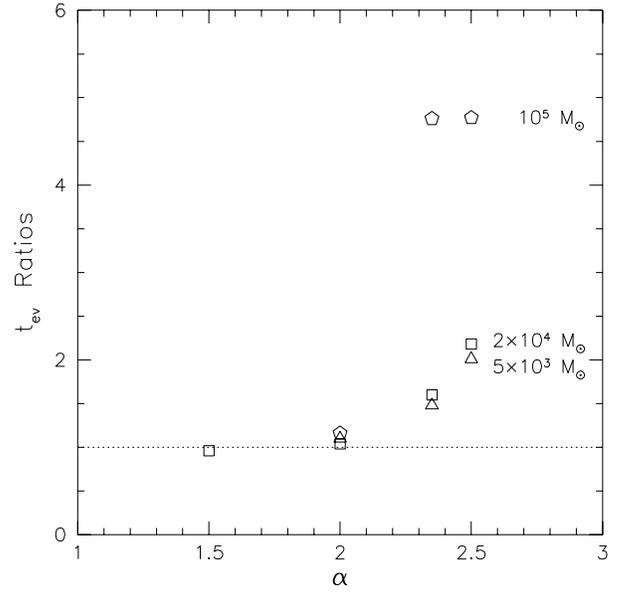}}
\caption
{\label{fig:ml}The ratios of $t_{ev}$ between models with the same parameters
except $m_l$ ($t_{ev}$ for $m_l=0.1 \msun$ over $t_{ev}$ for $m_l=1 \msun$).
Models shown here have $R_g=30$~pc and $W_0=4$.
Triangles are for $M=5 \times 10^3 \msun$, squares for $M=2 \times 10^4 \msun$,
and pentagons for $M=10^5 \msun$.  The dotted line represents $t_{ev}$ ratios
of unity.  For $\alpha \lsim 2$,
$t_{ev}$ is nearly independent of $m_l$ in range $0.1 < m_l / \msun < 1$.
For $M \lsim 2 \times 10^4 \msun$, $t_{ev}$ differs only by a factor of
$\lsim 2$ up to $\alpha=2.5$ in the same $m_l$ range.
}
\end{figure}

The postcollapse density profiles are not far from the profile at the
end of the collapse, and the latter may be described by the empirical relations:
\begin{mathletters}
\label{collapse}
\begin{eqnarray}
        \rho_i(0) & \propto & B(3/2,p_i+1)(-E_0)^{p_i+3/2}
                \label{collapse1} \\
        {d \ln \rho_i \over d \ln r} & = & -0.23(p_i+3/2)
                \label{collapse2} \\
        {p_i \over m_i} & = & {p_u \over m_u}
                \label{collapse3}
\end{eqnarray}
\end{mathletters}
where subscripts $i$ and $u$ are for the $i$th component and the most
massive component, respectively, $B(x,y)$ is the Beta function, and
$E_0$ the central potential (\markcite{BW77}Bahcall \& Wolf 1977,
\markcite{C85}Cohn 1985).  The above relation stems from $f_i \propto
(-E)^{p_i}$, which is an approximate distribution function that
a cluster has during the collapse.  \markcite{C85}Cohn found that
$p_u \approx 8.2$.  Equations (\ref{collapse1}) and (\ref{collapse2})
give nearly the same profiles for $p_i \ll 1$, which is seen also in
Figure \ref{fig:prof}.  Thus when a cluster has a wide mass range, its outer
region is dominated by heavier components, and
$t_{rh}$ is not dependent on $m_l$ as long as $m_l \ll m_u/8.2$
($\ll m_{15}/8.2$ in our case).  Such insensitivity of $t_{rh}$ to $m_l$
is also seen between models 113 \& 115 (smaller $\alpha$), models 111 \& 117
(smaller $W_0$), models 112 \& 118 (larger $W_0$), models 125 \& 129
(smaller $M$), and models 122 \& 127 (larger $M$).
On the other hand, $t_{ev}$'s of models with $\alpha \geq 2.35$ do depend
on $m_l$, especially for $M=5 \times 10^3 \msun$ (see Figure~\ref{fig:ml}).
This is because their mass ranges [$m_l$, $m_{15}$] are not wide enough.

\subsection{Initial Mass Function ($\alpha$)}
\label{sec:alpha}

\begin{figure}
\centerline{\epsfxsize=16cm\epsfbox{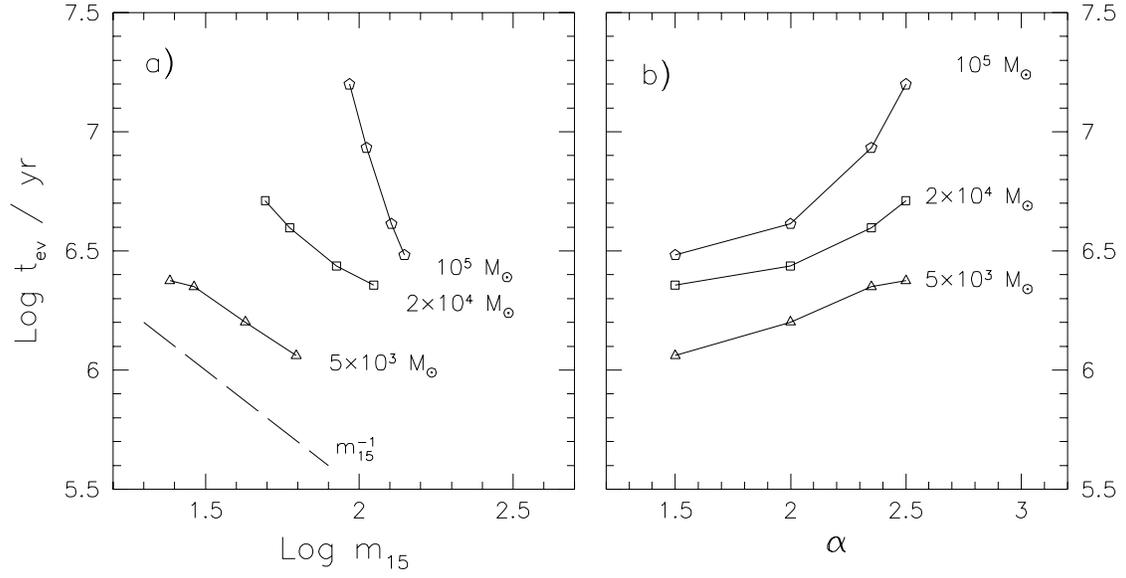}}
\caption
{\label{fig:alpha}$t_{ev}$ as a function of $m_{15}$ (a) and $\alpha$ (b)
for models with $m_l = 1 \msun$ and $R_g =30$~pc, and $W_0$=4.
Symbols are the same as in Figure~\ref{fig:ml}.
Dashed line represents $t_{ev} \propto m_{15}^{-1}$ relation.
}
\end{figure}

We discuss the effect of $\alpha$ on $t_{ev}$ in terms of $m_{15}$
(recall that $m_{15}$ is a function of $\alpha$).  Models 113, 101, 142,
and 114 have the same $M$ ($=2 \times 10^4 \msun$) but different
$\alpha$ values.  As noted above, $t_{ev}$ is inversely proportional
to $m_{15}$ (see Figure \ref{fig:alpha}) because of the predominance
of heavier components in determining $t_{rh}$ after rapid mass segregation.
Models 124, 125, 141, and 126 ($M=5 \times 10^3 \msun$) show the same
proportionality too.

Models 121, 122, 143, and 123 ($M=10^5 \msun$; different $\alpha$)
do not exactly follow the $t_{ev} \propto m_{15}^{-1}$ relation (although
there is still a clear inverse relation between $t_{ev}$ and $m_{15}$).
As a cluster's mass becomes larger, its lifetime and $m_{15}$ increase,
and therefore the stellar evolution timescale is a relatively smaller
fraction of the cluster's lifetime, and the relative amount of mass
that is subject to loss by stellar evolution before cluster evaporation
becomes larger.  Since the relative amount of mass that is subject to
loss by stellar evolution in a given period is a function of $\alpha$,
$t_{ev}$ is not simply proportional to $m_{15}^{-1}$ for the more massive
clusters in which the stellar evolution effect is not negligible.

\subsection{Galactocentric Radius ($R_g$)}
\label{sec:rg}

\begin{figure}
\centerline{\epsfxsize=16cm\epsfbox{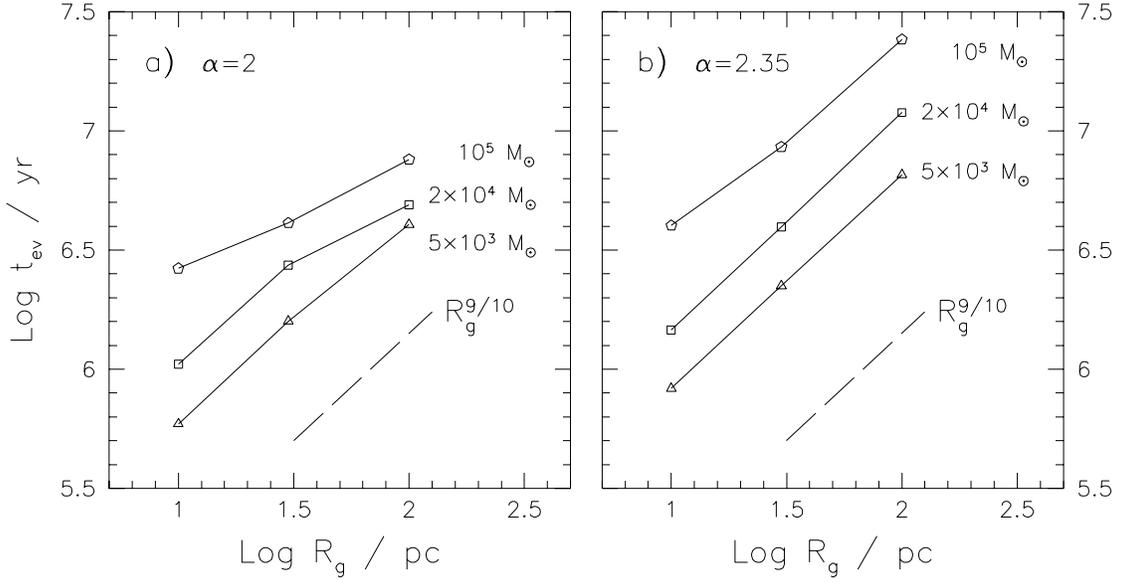}}
\caption
{\label{fig:rg}$t_{ev}$ as a function of $R_g$ for models with $m_l=1 \msun$,
$W_0=4$, and $\alpha=2$ (a) and 2.35 (b).
Symbols are the same as in Figure \ref{fig:ml}.  Without stellar evolution,
$t_{ev} \propto R_g^{9/10}$ (dashed lines) would be expected.
}
\end{figure}

Without stellar evolution, clusters with the same $M$
would evolve with a time scale proportional to $t_{rh} \propto R_t^{3/2}
\propto R_g^{9/10}$ (see eqs. [\ref{trh}], [\ref{rt}], and [\ref{mg}]).
However, the effect of stellar evolution on $t_{ev}$ is important in
many of our models, and the applicability of this relation is limited.
Figure~\ref{fig:rg} shows $t_{ev}$ of models with $\alpha=2$ and 2.35 as
a function of $R_g$ and $M$.
In case of $\alpha=2$, the $t_{ev} \propto R_g^{9/10}$ relation
holds only for the models with $M=5 \times 10^3 \msun$,
where models end before any significant stellar evolution, due to relatively
low $m_{15}$  values.  Stellar evolution significantly
affects $t_{ev}$ of the other $\alpha=2$ models, for which the exponent
in the above relation is smaller than 9/10.  For $\alpha=2.35$ models,
where the role of massive stars is relatively less important,
the effect of stellar evolution is much less apparent.  Models with
$M \leq 2 \times 10^4 \msun$ end before stellar evolution has
had a significant effect, and the
slope of $\log t_{ev}$ over $\log R_g$ of models with $M = 10^5 \msun$
is affected by stellar evolution only slightly.

\section{DISCUSSION}
\label{sec:discussion}

\subsection{Overview of the Results}

\begin{figure}
\centerline{\epsfxsize=8.8cm\epsfbox{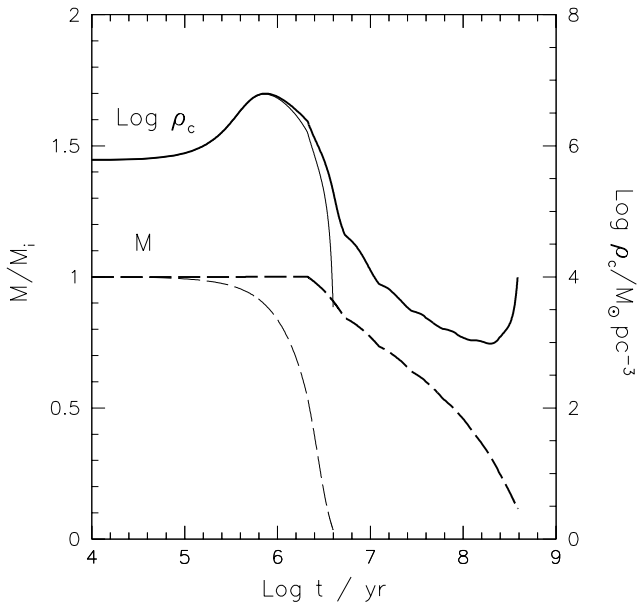}}
\caption
{\label{fig:tidal}$\rho_c$ (solid lines) and $M$ (dashed lines) evolution
of model 142 (thin lines) and the same model with a ten times larger $R_t$
(thick lines).  The density profile of the latter is set to be the same
as the former, so the cluster of the latter model does not initially fill
up the tidal radius.  This figure clearly shows the effects of strong
tidal fields on $t_{ev}$.  The $\rho_c$ of the model with a ten times
larger $R_t$ has a second core collapse at the end of the evolution.
}
\end{figure}

Our $t_{ev}$ survey results show that for our parameter regime, CYCs
will evaporate in less than or about 10~Myr, except for some cases with
$M=10^5 \msun$ {\it and} $\alpha \geq 2.35$.  The compactness of CYCs,
strong tidal fields, and large stellar mass ranges are jointly responsible
for the short lifetimes of CYCs, but among these, the strong tidal
fields play the most important role.  Figure \ref{fig:tidal} compares
the evolution of model 142 with a model which is the same in every respect,
except that $R_t$ is ten times larger (because the density profile is the
same as in model 142, the cluster does not initially fill up the tidal radius
in this case).  Here, the effect of the tidal field is dramatically illustrated:
when CYCs are located in tidal fields much weaker than in the GC, their
lifetimes are several orders of magnitude longer.  This could be the case
for the R136 cluster in the LMC, which has cluster parameters similar to
those of CYCs, but is apparently subjected to a much weaker tidal field.

\subsection{Comparison with Observations}

\begin{figure}
\centerline{\epsfxsize=8.8cm\epsfbox{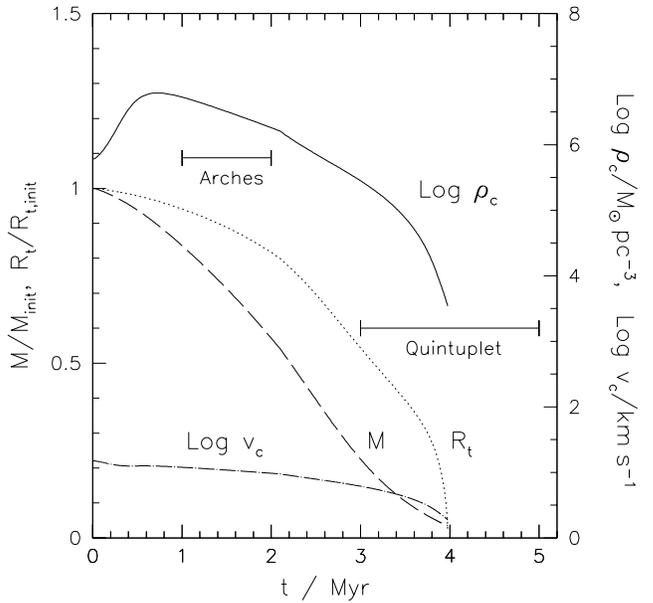}}
\caption
{\label{fig:evol}Evolution of $\rho_c$ (solid line), $M$ (dashed line),
$R_t$ (dotted line), and $v_c$ (dash-dotted line) of model 142.
$M$ and $R_t$ are normalized with their initial values.
The estimated locations in $t$-$\rho_{av}$ plane for the Arches
and Quintuplet clusters are indicated with bars (data from
Table~\ref{table:clusters}; note that $\rho_c$'s of the clusters
will be slightly higher than $\rho_{av}$ expecially for the Arches which
shows a high central concentration).  The Arches, the more
compact of the two, is situated near the epoch of core collapse, and
the Quintuplet, which looks much less bound, corresponds to the final
disruption phase of model 142.  The Figure may imply that two clusters
had similar initial conditions, and represent two distinctly different
epochs in their evolution.
}
\end{figure}

Of the Arches and Quintuplet clusters, the former is significantly
more compact, and is estimated to be younger and more massive
(see Table \ref{table:clusters}).  The smaller central density
of the Quintuplet
may be due to an initial condition with lower concentration or
smaller $M$ than that of the Arches.  However, because their ages
are probably quite different, it is
interesting to suppose that the two clusters had the same
initial conditions and to see if they represent two distinct
epochs in the evolutionary track of a cluster.
Since the Arches is estimated to have $M=1$-$2 \times 10^4 \msun$ and
$R_g \sim 30$~pc, our models with $M=2 \times 10^4 \msun$ and $R_g = 30$~pc
can be considered to represent these clusters.
Figure \ref{fig:evol} shows the evolution of model 142, whose parameters
are $M = 2 \times 10^4 \msun$, $\alpha = 2.35$, $W_0=4$,
$R_g=30$, and $m_l = 1 \msun$.  By comparison with the locations of the
clusters in this diagram, we infer that the Arches is located
near the epoch of core collapse, and the Quintuplet is near
the end of the evolution.
Figure \ref{fig:evol} lends plausibility to the idea that
the two clusters started with the same initial conditions and that
their largely different central densities are simply due to the age
difference.

The two clusters may equally well be represented as different evolutionary
stages of other model clusters having different values of $\alpha$ and $m_l$.
Model 114, which has the same parameters as model 142 except for
$\alpha =2.5$, gives an evolutionary route similar to that of model 142,
with a slightly longer $t_{ev}$.  Meanwhile, smaller $\alpha$
values result in shorter $t_{ev}$, and smaller $m_l$ values give longer
$t_{ev}$.  Thus clusters with an appropriate combination of smaller
$\alpha$ and $m_l$ will also be able to describe the two clusters together.

\begin{figure}
\centerline{\epsfxsize=8.8cm\epsfbox{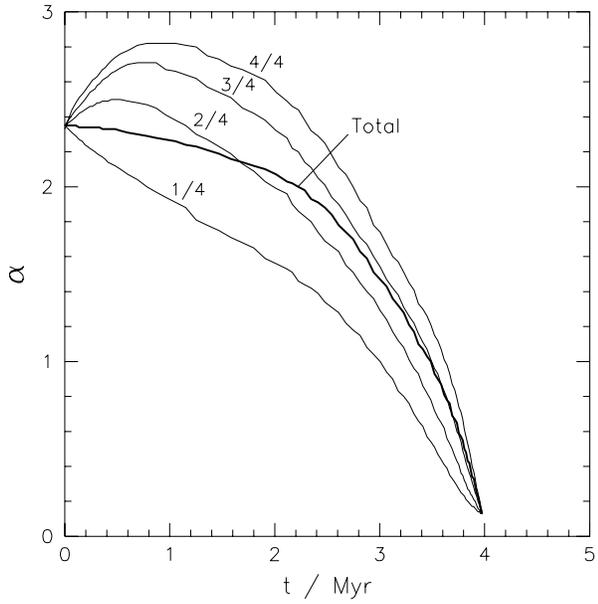}}
\caption
{\label{fig:mf}Evolution of $\alpha$ for the whole cluster (thick line)
and four inner-to-outer, equally-spaced annuli between the center and
$R_t$ (thin lines; from bottom to top) of model 142.  Considering the lower
detection limit in observations and stellar evolution of massive stars,
$\alpha$ was obtained for the mass range of [3,~30]$\msun$.
}
\end{figure}

The mass function evolves due to mass segregation, selective ejection
of lighter stars, and stellar evolution.  Furthermore, the mass function
varies with distance from the cluster center.  Thus observations of the
mass functions at several different radii, along with a determination of the
cluster age, would all be needed to infer the cluster IMF.   To illustrate
this point, we present in Figure \ref{fig:mf} the evolution of
$\alpha$ for the whole cluster and for four equally-spaced
annuli of model 142 .  Considering the observational inaccessibility of low
mass stars, and the rapid evolution of massive stars, the determination of
$\alpha$ was restricted to the mass range [3,~30]$\msun$.  While the
whole cluster mass function evolves relatively slowly until before
the final disintegration phase, the mass function of each annulus
evolves rather rapidly from the beginning.  The differences in $\alpha$
values between annuli are largest near the core collapse.  The
$\alpha$ values of the outer annuli initially grow due to mass
segregation, and later decrease as the tidal radius shrinks.  Mass
functions of CYCs measured with {\it HST} will soon be reported 
(\markcite{Fe99}Figer et al. 1999), and the radial dependence must
clearly be considered.  Since the mass function evolves
from the very beginning, even the current mass function of the younger
CYC, the Arches, will somewhat differ from the IMF.

The final disintegration of a cluster takes place in a relatively
short period of time compared to the cluster's lifetime.
\markcite{FH95}Fukushige \& Heggie (1995) argue that a cluster finally disrupts
by losing equilibrium when $R_h/R_t$ exceeds a certain critical value.
At this point, the central density and velocity dispersion decrease
abruptly, as may seen in Figure~\ref{fig:evol}.
We find that our models have $v_c$ less than or about $5 \, {\rm km \, s^{-1}}$
only during the final disruption phase, before which clusters maintain
$v_c$'s larger than or about $10 \, {\rm km \, s^{-1}}$.  Therefore,
an observation of $v_c$ for the Quintuplet will be able to provide
important information on the cluster's position in its
evolutionary track.

If the Quintuplet is indeed in a disruption phase,
$t_{ev}$ values calculated in this study may give
a constraint on the initial condition of the cluster:
the Quintuplet is not likely to initially have $\alpha \gsim 2.5$
if its $m_l$ is as low as $0.1 \msun$, because the Quintuplet
would not otherwise be in a disruption phase at its supposed current age.

\subsection{The Star Formation Rate}

The expected number of observed clusters at a certain epoch
in the inner bulge, $N_{obs}$, is basically a function
of $M$, $t_{ev}$, and the star formation rate (SFR) in the region:
\begin{equation}
\label{Nobs1}
        N_{obs} = \int \int {SFR \over M} t_{ev} (M, R_g)
                  \phi(M,R_g) \, dR_g \, dM,
\end{equation}
where $\phi (M,R_g)$ is a normalized probability that
a cluster is formed with $M$ at $R_g$.  We assume initially that stars are
all born in clusters.
As shown in Figure \ref{fig:rg}, $t_{ev}$ should apparently
be dependent on both $M$ and $R_g$.  Thus here we adopt a simple,
order-of-magnitude analysis for the relation between
$t_{ev}$ and $N_{obs}$.  First we split the variables in $t_{ev}$
and $\phi$.  For the former, we adopt a simple form,
\begin{equation}
        t_{ev} = 3 \times 10^6 \, {\rm yr} \, \left ( {R_g \over {\rm 30 \, pc}}
                 \right )^{0.75} \left ( {M \over 2 \times 10^4 \msun} \right ).
\end{equation}
The virtue of this approximation is that it, along with variable
splitting for $\phi$, makes equation (\ref{Nobs1}) independent of $M$,
whose distribution is not well known (in fact, the behaviour of $t_{ev}$'s
is better described with $\propto M^{1/2}$ rather than $\propto M$,
but we choose the latter for simpler estimation).  For the latter,
we define $\phi \equiv \phi_M(M) \phi_R(R_g)$, where $\phi_M$
and $\phi_R$ are normalized for their variables, and we assume
that $\phi_R$ follows the density profile in the inner bulge $\rho_g$:
$\phi_R \propto R_g^2 \rho_g \propto {\rm const}$ ($\rho_g \propto R_g^{-2}$
is adopted instead of $\rho_g \propto R_g^{-1.8}$ for simplicity).
We take [0,~100]~pc for the range of $R_g$
(the final result is not so sensitive to the choice of this range).
Then equation (\ref{Nobs1}) now becomes
\begin{eqnarray}
\label{Nobs2}
        N_{obs} & \simeq & 15 \left ( {SFR \over 0.1 \msun \, {\rm yr^{-1}}}
                           \right ) \int_{0 \, {\rm pc}}^{100 \, {\rm pc}}
                           \left ( {R_g \over 30 \, {\rm pc}}
                           \right )^{0.75} \nonumber \\
                &        & \; \; \; \; \; \; \; \; \; \; \times \phi_R(R_g) \,
			   dR_g \; \int \phi_M(M) \, dM \nonumber\\
                & \simeq & 20 \left ( {SFR \over 0.1 \msun \, {\rm yr^{-1}}}
                           \right ).
\end{eqnarray}
$N_{obs}$ is then equal to the actual current
number of large, distinguishable clusters in the inner bulge, two,
if SFR is as low as $0.01 \msun \, {\rm yr^{-1}}$, a value which happens
to be very close to the mass of the two
clusters divided by their lifetime, which is a crude estimate
of the SFR without any consideration of dependencies on $R_g$ and $M$.

\markcite{G89}G\"usten (1989) estimated the SFR in the inner bulge to be
0.3-0.6~$M_\odot \, {\rm yr^{-1}}$ from the global production rate of
Lyman continuum photons, and 0.05~$M_\odot \, {\rm yr^{-1}}$
from the luminosity of the discrete far-IR sources measured
by \markcite{OF84}Odenwald \& Fazio (1984).
G\"usten's estimates are based on the assumption
of a Salpeter mass function with $m_l$ of $0.1 \msun$.  For a flatter
mass function with elevated $m_l$, his SFR becomes smaller, although
still considerably larger than 0.01~$M_\odot \, {\rm yr^{-1}}$.  It thus
appears that a considerable fraction
of stars in the inner bulge forms outside of clusters, or in much smaller
clusters than the Arches and Quintuplet.

A number of other sites of star formation are known in the
GC, but none apparently has a stellar luminosity
within an order of magnitude of the Arches and Quintuplet.
Partial surveys have been carried out at near-IR wavelengths
to identify other potential clusters near the GC
(\markcite{CWG90}Catchpole, Whitelock, \& Glass 1990; \markcite{F95}Figer 1995;
\markcite{Pe99}Philipp et al. 1999), but no others have yet been
identified, and it seems likely that if there are any
others comparable to the Arches or Quintuplet, they are
highly obscured, even at 2 microns.

We therefore speculate that the circumstances of the formation
of these clusters was peculiar in some fundamental way, such
as catastrophic formation by extremely strong shocks, or direct
collision of two dense molecular clouds.  If so, then young stars near
the GC are formed in multiple modes, and one
cannot ascribe the total star formation rate there to massive
clusters alone.

\subsection{Binary Heating and Close Encounters}

We find that in all of our models, heating by three-body binaries
exceeds that by tidal-capture binaries during the core collapse, and that
the postcollapse expansion is driven by three-body binaries.
Moreover, in most models, the cumulative number of
tidal binaries formed is less than unity until the end of evolution.
These numbers are one to two orders of magnitude smaller than the number of
collisional mergers between stars during the first few Myr
calculated by \markcite{PZe99}Portegies Zwart et al. (1999) for the R136
cluster in the LMC with initial central mass density similar to our models.
The discrepancy is due to the fact that we did not consider
the evolution of massive stars into a giant phase in which the stellar radii
are largely increased.  Thus our tidal-capture binary formation rates are
somewhat underestimated, but since the heating per binary is much larger
for three-body binaries, the inclusion of radius evolution into our
calculations would not change the overall dynamical evolution of the clusters
except only for some cases with considerably large initial central densities
and steep IMFs (these conditions are more favorable to tidal-capture
binary formation; see, e.g., \markcite{KLG98}Kim, Lee, \& Goodman 1998).
In these cases, the postcollapse expansion will be driven by tidal-capture
binaries and clusters will expand more slowly than the ones driven by
three-body binaries.

\subsection{Dynamical Friction}

Clusters are subject to dynamical friction as they go through
field stars while orbiting around the galaxy center.
The drag by dynamical friction causes the clusters to lose
energy and spiral in toward the center, where the tidal field is
stronger.  When the field stars
have a density distribution $\propto R_g^{-2}$, the time
required for a cluster initially on a circular orbit to
reach the center is given by
\begin{eqnarray}
\label{tfric}
        t_{fric} \simeq {2 \times 10^9 \, {\rm yr} \over
                        \ln \Lambda } \left ( {R_g \over
                        30 \, {\rm pc}} \right )^2
                        \left ( {M \over 2 \times 10^4 \msun}
                        \right )^{-1} \nonumber \\
			\hfill \left ( {v_c \over 170 \,
                        {\rm km \, s^{-1}}} \right ),
\end{eqnarray}
where $v_c$ is the circular orbital velocity of the cluster and
$\ln \Lambda$ is the Coulomb logarithm for the field stars
(\markcite{BT87}Binney \& Tremaine 1987).
Reasonable values for the inner bulge are $v_c \simeq 170 \,
{\rm km \, s^{-1}}$ and $\ln \Lambda \approx 10$.  Then
$t_{fric}$ would be comparable to or smaller than $t_{ev}$
calculated in this study only for clusters with extremely small
$R_g$ and/or large $M$.  Thus, most CYCs will disintegrate
through tidal evaporation before a significant orbital decay
due to dynamical friction takes place.

\subsection{Limitations}

We assumed that clusters consist of coevally formed stars and
start their main-sequence stages at the beginning of simulations.
However, stars may gravitationally interact with others already
from pre-main-sequence or even from accreting protostar stages
and these stages may be important to CYCs because of their short $t_{ev}$'s.
For very massive stars, the time scale of these stages prior to
the main-sequence must be much shorter than $t_{ev}$'s calculated
in this study, but it becomes larger than 10~Myr for stars lighter
than $2 \msun$ (\markcite{B96}Bernasconi 1996).  Thus if lighter stars
systematically form earlier than heavier stars, the dynamical evolution of
lighter stars until the formation of heavier stars may become important.
The role of the stages prior to main-sequence
on the cluster's lifetime is not clear though, because not much
is yet known about the details of cluster formation, such as
what fraction of the gas goes into star formation (star formation
efficiency), how quickly the gas left over from star formation
disappears from the cluster by stellar wind and supernova
explosions (timescale of residual gas expulsion), and which gas
clumps (as a function of mass) contract into stars first and
act as discrete gravity sources.
These uncertainties are particularly large for the unusual
clusters we are considering here.
These issues are in the arena
of N-body simulations and the first two issues were considered
by \markcite{TTe86}Tenorio-Tagle et al. (1986) and
\markcite{G97a}\markcite{G97b}Goodwin (1997a,b) for young
globular clusters.  No observational evidence for the presence
of residual gas in the Arches and Quintuplet has been found yet.
Proper inclusion of these effects
into early cluster dynamics will be possible only after a significant
advance in the study of star formation is achieved.

Throughout the calculations, we have set $N_{15} = 50$, but this value
is arbitrarily chosen.  Higher values would give more statistically
correct results, but would not represent the whole mass range well.
This is one of the intrinsic limitations with Fokker-Planck simulations.
To see the dependence of $t_{ev}$ on $N_{15}$, we performed models 142 and
145 with $N_{15} = 150$, and obtained only $\sim 25$~\% larger $t_{ev}$ values.
This implies that the choice of $N_{15}$ does not significantly alter our
general results.  On the other hand, our single-power-law IMFs may
be too simple to represent realistic IMFs.  Thus we tried a Kroupa IMF
(\markcite{KTG93}Kroupa, Tout, \& Gilmore 1993), which is a three-part
power-law with $\alpha = 2.7$ for $m > 1 \msun$, 2.2 for $0.5 \msun < m <
1 \msun$, and 1.3 for $0.08 \msun < m < 0.5 \msun$, for $M = 2 \times 10^4
\msun$, $R_g = 30 \, {\rm pc}$, and $W_0 = 4$.  We found that the model with
a Kroupa IMF has results similar to those of model 116, in which the parameters
are $\alpha = 2.5$ and $m_l = 0.1 \msun$.  Since the Kroupa IMF was derived
for the Galactic disk, it may not be applicable to the GC where the star
formation environment is largely different from that in the disk, but
we believe that even the evolution of a cluster with realistic GC IMF
can be well approximated with one of our simple power-law IMFs.

Orbit-averages used in our Fokker-Planck simulations are valid when
$t_{dyn} \ll t_{rh}$.
Since $t_{rh} \approx (0.1 N/ \ln \Lambda) t_{dyn}$, CYCs, which have $N$
of $\sim 10^{3-5}$, apparently satisfy the requirement.  However, the wide
mass range of CYCs shortens the relaxation time in the core through
mass segregation, and the above requirement may not hold in the core
for some of our models, especially the ones with the largest mass ranges.
In addition, while the Fokker-Planck method also assumes $t_{dyn} \ll
t_{se}$, some of our models with large $M$ and small $\alpha$
may have periods when the assumption is temporarily violated.  The
validity of using the Fokker-Planck method for these marginal situations
can be assessed using N-body simulations.

\section{SUMMARY}
\label{sec:summary}

We have investigated the dynamical evolution of CYCs near the
Galactic center with anisotropic Fokker-Planck models.  Stellar
evolution and heating by three-body and tidal-capture binaries
were included in the calculation, and the apocenter criterion with
an appropriate removal speed constant was adopted for the tidal
evaporation.  Mass bins were chosen depending on cluster parameters
to properly account for a relatively small number of most massive
stars.

The evolutionary time scales of CYCs are fairly short because of
their compactness and strong tidal fields.  For our parameter regime
($1.5 \leq \alpha \leq 2.5$, $m_l=0.1$~\&~1$\msun$, $5 \times 10^3 \msun
\leq M \leq 10^5 \msun$, and $10 \, {\rm pc} \leq R_g \leq 100 \, {\rm pc}$),
clusters evaporate in $\lsim 10$~Myr except for few clusters
with $M=10^5 \msun$.  Core collapse takes place in most clusters, and
three-body binaries dominate the heating.

Unlike globular clusters, very massive stars in CYCs play important
roles in dynamics because relaxation and mass segregation times are
comparable to or even smaller than those stars' lifetimes.
Rapid mass segregation due to a large difference
in upper and lower mass boundaries accelerates a cluster's evolution even
more.  Strongly selective ejection of lighter stars makes massive stars
dominate even in the outer regions of the cluster and the
evolution of such clusters is weakly dependent on the lower mass boundary.

We found that the Arches and Quintuplet clusters, which have quite
different central densities, central concentrations, and ages,
may be placed on one evolutionary track.  Among our cluster models,
that with $M=2 \times 10^4 \msun$, $R_g =30$~pc, $\alpha=2.35$ or 2.5,
and $m_l=1 \msun$ gives the evolutionary track that best describes both
clusters simultaneously.  The first two parameters are from observational
constraints, and an appropriate combination of smaller $\alpha$ and $m_l$
can also describe the two clusters together.

The total mass in observed CYCs is several $10^4 \msun$ and the lifetime
of CYCs with observed $M$ at observed $R_g$ is a few Myr.  These numbers
result in an inner bulge SFR of $\sim 0.01 \msun \, {\rm yr^{-1}}$.
A more detailed (but still approximate) estimate of the SFR with a
consideration of distributions in $M$ and $R_g$ happens to be similar
to this simple estimate.  Lyman continuum photons and far-IR
observations in the inner bulge region give SFRs of
0.05--0.5$\msun \, {\rm yr^{-1}}$.
We suggest that the mode of formation of CYCs is peculiar, likely resulting
from strong shocks or collisions of dense molecular clouds.

\acknowledgements
S.S.K. is deeply grateful to Koji Takahashi for generously providing us
of his anisotropic Fokker-Planck codes and for his kind help with the codes.
We thank Don Figer, Cheongho Han, Eunhyeuk Kim, Pavel Kroupa, Myung Gyoon Lee,
Kap Soo Oh, R. Michael Rich, and Rainer Spurzem for helpful discussions.
This work was supported in part by the International Cooperative Research
Program of the Korea Research Foundation to Pusan National University in 1998,
and in part by NASA through grant number GO-07364.01-96A to UCLA from the Space
Telescope Science Institute, which is operated by AURA, Inc., under NASA
contract NAS5-26555.


\clearpage

\begin{deluxetable}{clccrcccrrr}
\tablecolumns{11}
\tablewidth{0pt}
\tablecaption{Simulation Parameters and Results
\label{table:runs}}
\tablehead{
\colhead{} &
\colhead{} &
\colhead{} &
\colhead{$M$} &
\colhead{$R_g$} &
\colhead{$R_t$} &
\colhead{$m_l-m_u$} &
\colhead{} &
\colhead{$\bar m$} &
\colhead{$m_{15}$} &
\colhead{$t_{ev}$} \\
\colhead{Models} &
\colhead{$\alpha$} &
\colhead{$W_0$} &
\colhead{($M_\odot$)} &
\colhead{(pc)} &
\colhead{(pc)} &
\colhead{($M_\odot$)} &
\colhead{$\beta$} &
\colhead{($M_\odot$)} &
\colhead{($M_\odot$)} &
\colhead{(Myr)}
}
\startdata
  101&        2&        4& $2\times 10^4$& 30&    1.11 &    1-150&    0.526&    5.04 &    84.3 &     2.7 \nl
  102&        2&        4& $2\times 10^4$& 30&    1.11 &  0.1-150&    0.722&    0.73 &    72.1 &     2.9 \nl
  103&        2&        4& $2\times 10^4$& 30&    1.11 &    1-75 &    0.659&    4.37 &    54.9 &     4.0 \nl
  111&        2&        1& $2\times 10^4$& 30&    1.11 &    1-150&    0.526&    5.04 &    84.3 &     2.7 \nl
  112&        2&        7& $2\times 10^4$& 30&    1.11 &    1-150&    0.526&    5.04 &    84.3 &     2.7 \nl
  113&      1.5&        4& $2\times 10^4$& 30&    1.11 &    1-150&    0.266&    12.25&    111.6&     2.3 \nl
  114&      2.5&        4& $2\times 10^4$& 30&    1.11 &    1-150&    0.763&    2.76 &    49.5 &     5.1 \nl
  115&      1.5&        4& $2\times 10^4$& 30&    1.11 &  0.1-150&    0.430&    3.87 &    109.9&     2.2 \nl
  116&      2.5&        4& $2\times 10^4$& 30&    1.11 &  0.1-150&    1.000&    0.29 &    26.7 &    11.2 \nl
  117&        2&        1& $2\times 10^4$& 30&    1.11 &  0.1-150&    0.722&    0.73 &    72.1 &     2.8 \nl
  118&        2&        7& $2\times 10^4$& 30&    1.11 &  0.1-150&    0.722&    0.73 &    72.1 &     2.8 \nl
  121&      1.5&        4& $1\times 10^5$& 30&    1.89 &    1-150&    0.966&    12.25&    140.3&     3.0 \nl
  122&        2&        4& $1\times 10^5$& 30&    1.89 &    1-150&    1.020&    5.04 &    127.3&     4.1 \nl
  123&      2.5&        4& $1\times 10^5$& 30&    1.89 &    1-150&    1.124&    2.76 &    92.9 &    15.8 \nl
  124\tablenotemark{a}&      1.5&        4& $5\times 10^3$& 30&    0.70 &    1-150&  \nodata&    12.25&    62.4 &     1.2 \nl
  125&        2&        4& $5\times 10^3$& 30&    0.70 &    1-150&    0.108&    5.04 &    42.7 &     1.6 \nl
  126&      2.5&        4& $5\times 10^3$& 30&    0.70 &    1-150&    0.384&    2.76 &    24.2 &     2.4 \nl
  127&        2&        4& $1\times 10^5$& 30&    1.89 &  0.1-150&    1.032&    0.73 &    119.5&     4.8 \nl
  128&      2.5&        4& $1\times 10^5$& 30&    1.89 &  0.1-150&    1.184&    0.29 &    59.3 &    75.5 \nl
  129&        2&        4& $5\times 10^3$& 30&    0.70 &  0.1-150&    0.506&    0.73 &    33.9 &     1.8 \nl
  130&      2.5&        4& $5\times 10^3$& 30&    0.70 &  0.1-150&    0.828&    0.29 &    12.0 &     4.8 \nl
  131&        2&        4& $5\times 10^3$& 10&    0.36 &    1-150&    0.108&    5.04 &    42.7 &     0.6 \nl
  132&        2&        4& $2\times 10^4$& 10&    0.57 &    1-150&    0.526&    5.04 &    84.3 &     1.1 \nl
  133&        2&        4& $1\times 10^5$& 10&    0.98 &    1-150&    1.020&    5.04 &    127.3&     2.7 \nl
  134&        2&        4& $5\times 10^3$&100&    1.43 &    1-150&    0.108&    5.04 &    42.7 &     4.1 \nl
  135&        2&        4& $2\times 10^4$&100&    2.28 &    1-150&    0.526&    5.04 &    84.3 &     4.9 \nl
  136&        2&        4& $1\times 10^5$&100&    3.89 &    1-150&    1.020&    5.04 &    127.3&     7.6 \nl
  141&     2.35&        4& $5\times 10^3$& 30&    0.70 &    1-150&    0.306&    3.19 &    29.0 &     2.2 \nl
  142&     2.35&        4& $2\times 10^4$& 30&    1.11 &    1-150&    0.691&    3.19 &    59.4 &     4.0 \nl
  143&     2.35&        4& $1\times 10^5$& 30&    1.89 &    1-150&    1.083&    3.19 &    105.6&     8.6 \nl
  144&     2.35&        4& $5\times 10^3$& 30&    0.70 &  0.1-150&    0.735&    0.36 &    16.9 &     3.3 \nl
  145&     2.35&        4& $2\times 10^4$& 30&    1.11 &  0.1-150&    0.913&    0.36 &    37.6 &     6.4 \nl
  146&     2.35&        4& $1\times 10^5$& 30&    1.89 &  0.1-150&    1.122&    0.36 &    79.3 &    40.8 \nl
  151&     2.35&        4& $5\times 10^3$& 10&    0.36 &    1-150&    0.306&    3.19 &    29.0 &     0.8 \nl
  152&     2.35&        4& $2\times 10^4$& 10&    0.57 &    1-150&    0.691&    3.19 &    59.4 &     1.5 \nl
  153&     2.35&        4& $1\times 10^5$& 10&    0.98 &    1-150&    1.083&    3.19 &   105.6 &     4.0 \nl
  154&     2.35&        4& $5\times 10^3$&100&    1.43 &    1-150&    0.306&    3.19 &    29.0 &     6.5 \nl
  155&     2.35&        4& $2\times 10^4$&100&    2.28 &    1-150&    0.691&    3.19 &    59.4 &    12.0 \nl
  156&     2.35&        4& $1\times 10^5$&100&    3.89 &    1-150&    1.083&    3.19 &   105.6 &    24.2 \nl
\tablenotetext{a}{$\beta$ becomes a negative value for this model, so an equal
logarithmic binning was used instead.}
\enddata
\end{deluxetable}

\begin{deluxetable}{lcccc}
\tablecolumns{5}
\tablewidth{0pt}
\tablecaption{Properties of Compact Young Clusters
\label{table:clusters}}
\tablehead{
\colhead{} &
\colhead{Log $M$} &
\colhead{$R_{av}$} &
\colhead{Log $\rho_{av}$} &
\colhead{Age} \\
\colhead{Cluster} &
\colhead{($M_\odot$)} &
\colhead{(pc)} &
\colhead{($M_\odot \, {\rm pc^{-3}}$)} &
\colhead{(Myr)}
}
\startdata
Arches     &  4.3  &  0.2  &  5.8  &  1-2   \nl
Quintuplet &  3.8  &  1.0  &  3.2  &  3-5   \nl
R136       &  4.5  &  1.6  &  3.3  & $<1-2$ \nl
\enddata
\tablecomments{From Table~5 of \markcite{FMM99}Figer et al. (1999).
$M$ is the total cluster mass in all stars extrapolated from observation
down to a lower-mass cutoff of $1 \msun$, assuming a Salpeter IMF slope
and an upper mass cutoff of $120 \msun$.  $R_{av}$ is the average
projected separation from the controid position.  Since the low-mass end
has not been identified in these clusters yet, the radius may be larger
than the value given here when mass segregation is present.  $\rho_{av}$
is $M$ divided by the volume inside $R_{av}$.}
\end{deluxetable}

\clearpage

\onecolumn

\twocolumn

\end{document}